\documentclass[prl,twocolumn,nofootinbib,nobibnotes,noeprint,preprintnumbers,floatfix]{revtex4-1}
\usepackage{amssymb,amsmath,graphicx}
\usepackage[dvipsnames]{xcolor}
\usepackage[colorlinks,allcolors=NavyBlue]{hyperref}
\usepackage{microtype}

\newcommand{\be}{\begin{eqnarray}}
\newcommand{\ee}{\end{eqnarray}}

\usepackage[normalem]{ulem}

\def\lsim{\mathrel{\raise.3ex\hbox{$<$\kern-.75em\lower1ex\hbox{$\sim$}}}}
\def\gsim{\mathrel{\raise.3ex\hbox{$>$\kern-.75em\lower1ex\hbox{$\sim$}}}}

\begin{document}

\preprint{FERMILAB-PUB-23-232-T}

\title{A Leptonic Model for Neutrino Emission From Active Galactic Nuclei}

\author{Dan Hooper$^{1,2,3}$}
\thanks{dhooper@fnal.gov, http://orcid.org/0000-0001-8837-4127}

\author{Kathryn Plant$^4$}
\thanks{kp@astro.caltech.edu, http://orcid.org/0000-0001-6360-6972}

\affiliation{$^1$Fermi National Accelerator Laboratory, Theoretical Astrophysics Department, Batavia, IL, USA}
\affiliation{$^2$University of Chicago, Department of Astronomy \& Astrophysics, Chicago, USA}
\affiliation{$^3$University of Chicago, Kavli Institute for Cosmological Physics, Chicago, IL, USA}
\affiliation{$^4$California Institute of Technology, Department of Astronomy, Pasadena, CA, USA}

\date{\today}

\begin{abstract}

It is often stated that the observation of high-energy neutrinos from an astrophysical source would constitute a smoking gun for the acceleration of hadronic cosmic rays. Here, we point out that there exists a purely leptonic mechanism to produce TeV-scale neutrinos in astrophysical environments. In particular, very high-energy synchrotron photons can scatter with X-rays, exceeding the threshold for muon-antimuon pair production. When these muons decay, they produce neutrinos without any cosmic-ray protons or nuclei being involved. In order for this mechanism to be efficient, the source in question must produce very high-energy photons which interact in an environment that is dominated by keV-scale radiation.
As an example, we consider the active galaxy NGC 1068, which IceCube has recently detected as a source of TeV-scale neutrinos. We find that the neutrino emission observed from this source could potentially be generated through muon pair production for reasonable choices of physical parameters.

\end{abstract}

\maketitle

\section{Introduction}

The conventional wisdom in the field of neutrino astrophysics is that the detection of high-energy neutrinos from a given source would be unambiguous evidence that that object accelerates cosmic-ray protons or nuclei. In particular, whereas gamma rays can be generated through both leptonic (inverse Compton scattering, synchrotron) and hadronic (pion production) processes, it has long been thought that high-energy astrophysical neutrinos would be produced only through the production and decay of pions, which are generated through the interactions of high-energy protons with gas or radiation. In this sense, neutrinos play a critical role in our efforts to identify the sources of the hadronic cosmic-ray spectrum.

If the diffuse spectrum of high-energy astrophysical neutrinos observed by IceCube~\cite{IceCube:2020acn,IceCube:2021rpz,IceCube:2013cdw,IceCube:2013low,IceCube:2014stg} is generated through hadronic interactions in optically thin environments ({\it i.e.}, those transparent to gamma rays), these neutrinos will inevitably be accompanied by gamma rays from the decays of neutral pions. Normalizing the pion production rate to the spectrum of neutrinos reported by IceCube, one finds that these sources should collectively generate a flux of gamma rays that would approximately saturate, or even exceed, the isotropic background reported by the Fermi Collaboration~\cite{Hooper:2016jls,Murase:2015xka,Giacinti:2015pya,Murase:2013rfa}. When this fact is combined with the lack of observed correlations between the directions of IceCube's neutrinos and known gamma-ray sources~\cite{IceCube:2019cia,IceCube:2016tpw,IceCube:2018omy,IceCube:2016ipa,Smith:2020oac,IceCube:2016qvd,Hooper:2018wyk}, transparent source scenarios appear to be somewhat disfavored, instead suggesting that many of these neutrinos are produced in optically thick environments, in what are known as ``hidden sources.'' From this perspective, the dense cores of Active Galactic Nuclei (AGN) are seen as a particularly well-motivated class of sources for IceCube's diffuse neutrino flux~\cite{Murase:2015xka,Khiali:2015tfa,Stecker:2013fxa,Kimura:2014jba,Kalashev:2014vya} (for a review, see Ref.~\cite{Murase:2015ndr}).

The IceCube Collaboration has recently reported an excess of 79 events from the direction of the nearby active galaxy, NGC 1068 (also known as M77), corresponding to a 4.2$\sigma$ detection of $\sim 1-10 \, {\rm TeV}$ neutrinos~\cite{IceCube:2022der} (see also, Ref.~\cite{IceCube:2019cia}). Although NGC 1068 has been detected by the Fermi telescope at $\sim 0.1-30 \, {\rm GeV}$ energies~\cite{Fermi-LAT:2019yla,Fermi-LAT:2019pir}, MAGIC has failed to detect gamma rays from source, placing strong limits on its emission in the $\sim 0.1-10 \, {\rm TeV}$ band~\cite{MAGIC:2019fvw}. The lack of TeV-scale gamma rays from this source allows us to rule out the possibility that the observed neutrinos are produced in an optically thin environment, instead favoring scenarios in which cosmic-ray protons are accelerated and produce pions in the dense region immediately surrounding this AGN's supermassive black hole~\cite{Murase:2022dog}. Observations by NuSTAR~\cite{Marinucci:2015fqo} and XMM-Newton~\cite{Bauer:2014rla} have detected bright X-ray emission from this source (corresponding to an intrinsic luminosity of $L_X  \sim 10^{44} \, {\rm erg/s}$ in the 2-10 keV band, and extending up to energies of $\epsilon_X \sim 10^2 \, {\rm keV}$), suggesting that the densities of high-energy radiation in the central region ({\it i.e.}, the corona) could be large enough to efficiently absorb gamma rays through pair production, while still allowing neutrinos to escape.

The conventional wisdom is that the neutrinos observed from NGC 1068 should allow us to definitively identify this object as an accelerator of cosmic ray protons. In this letter, we propose an alternative mechanism for generating the neutrino emission from AGN which is purely leptonic in nature. In particular, cosmic ray electrons in or near the AGN's corona could produce very high-energy gamma rays which would scatter with X-rays to produce muon-antimuon pairs. These muons would then decay to produce neutrinos, without any need for high-energy protons. No protons would be harmed in the making of these neutrinos.

\section{Neutrinos from Muon Pair Production}

The production of very high-energy photons through the process of synchrotron emission requires the presence of very high-energy electrons in a region with a very strong magnetic field~\cite{Blumenthal:1970gc}. Equating the timescales for acceleration and synchrotron losses, the maximum energy to which an electron can be accelerated is given by (see, for example, Ref.~\cite{Sudoh:2019jup})
\begin{align}
\label{eq:Emax}
E_e^{\rm max} \sim 300 \, {\rm TeV} \times \bigg(\frac{B}{0.03 \, {\rm G}}\bigg)^{-1/2},
\end{align}
where $B$ is the strength of the magnetic field.
%
The intensity of the synchrotron radiation from a relativistic electron peaks near the critical frequency, $\nu_c$, corresponding to an energy of
\begin{align}
\label{eq:Ecrit}
E_{\rm syn} &\sim h \nu_c =  \frac{3\pi E_e^2 \nu_g \sin \alpha_p}{m^2_e}  \\
&\approx 9 \, {\rm TeV} \times   \bigg(\frac{E_e}{300 \, {\rm TeV}}\bigg)^2\,\bigg(\frac{B}{2 \times 10^3 \, {\rm G}}\bigg) \, \bigg(\frac{\sin \alpha_p}{\sqrt{2/3}\,}\bigg), \nonumber 
\end{align}
where $\nu_g = eB/2\pi m_e$ is the the non-relativistic gyrofrequency, and $\alpha_p$ is the pitch angle. In combining Eqs.~\ref{eq:Emax} and~\ref{eq:Ecrit}, we find that for the case of a uniform magnetic field, synchrotron photons are limited to energies below $E_{\rm syn} \sim 0.1 \, {\rm GeV}$, corresponding to what is known as the ``burnoff limit''~\cite{1996ApJ...457..253D}. As we will show below, the production of neutrinos through muon pair production requires $\gsim \mathcal{O}(1-10 \, {\rm TeV})$ photons, well above what is allowed by the burnoff limit.
Synchrotron emission, however, could potentially reach these energies in scenarios in which electrons are accelerated in regions with relatively small magnetic fields (such as in the torus, for example~\cite{Lopez_Rodriguez_2020}, or in shocks in the corona~\cite{Inoue:2019yfs,Inoue:2019fil}) before they encounter regions of stronger magnetic fields, possibly in dense clumps within the corona~\cite{Khangulyan:2020zdu}.
%
%
%
We could also consider scenarios that feature anisotropic acceleration, or synchrotron emission that is produced from a population of electrons with a significant bulk Lorentz factor. 

Alternatively, the very high-energy photons that are required in this scenario could be produced through the process of inverse Compton scattering. In particular, for target photons with energies near $\epsilon \sim m^2_e/E_{e} \sim 0.03 \, {\rm eV} \times (10 \, {\rm TeV}/E_e)$, such scattering would occur near the boundary of the Thomson and Klein-Nishina regimes, yielding $E_{\gamma} \sim E_e$, but without suffering from a large degree of Klein-Nishina suppression. Note that if the very high-energy photons are produced through inverse Compton scattering, no large magnetic fields would be required.

To estimate the spectrum of the synchrotron radiation that is emitted from a population of electrons, we adopt the simplifying approximation that each electron radiates the entirety of its power at its critical frequency. While this is not precisely correct, the power does climb until $\nu \sim \nu_c$, and falls exponentially at higher frequencies, leading to results that are not very different from those that would be obtained in a more careful calculation. Under this approximation, the total synchrotron power per unit frequency can be written as follows:
\begin{align}
j(\nu) \approx \bigg(\frac{dN_e}{dE_e}\bigg) \, \bigg(\frac{\Delta E_e}{\Delta \nu_c}\bigg) \, \bigg(\frac{-dE_e}{dt}\bigg), \,
\end{align}
where $dN_e/dE_e$ is the spectrum of the radiating electrons, $\Delta E_e/\Delta \nu_c = E_e/2 \nu_c$ is the degree to which the critical frequency of the synchrotron emission changes as an electron cools, and $-dE_e/dt$ is the synchtron energy loss rate. After averaging over the pitch angle, the energy loss rate from synchrotron is given by~\cite{Blumenthal:1970gc}
\begin{align}
\label{eq:syncloss}
\frac{dE_e}{dt} &= - \frac{2 \sigma_t B^2 E^2_e}{3  \mu_0 m^2_e}  \\
&\approx -2.5 \times 10^3 \, {\rm TeV}/{\rm s} \times \bigg(\frac{E_e}{{\rm TeV}}\bigg)^2 \, \bigg(\frac{B}{10^3 \, {\rm G}}\bigg)^2, \nonumber
\end{align}
where $\sigma_t$ is again the Thomson cross section

For the case of a power-law spectrum of electrons, $dN_e/dE_e = A E_e^{-p}$, the spectrum of synchrotron emission is approximately given by
\begin{align}
\label{eq:syncpowerlaw}
j(\nu) &\approx  \bigg(AE_e^{-p}\bigg) \, \bigg(\frac{E_e}{2\nu}\bigg) \, \bigg(\frac{2\sigma_t B^2 E^2_e}{3 \mu_0 m^2_e}\bigg) \\[10pt]
&\approx \frac{A\sigma_t B^2}{3\mu_0 m_e^2 \nu} \, \bigg(\frac{2\pi m_e^3\nu}{eB}\bigg)^{(3-p)/2}, \nonumber 
\end{align}
where, in the second line, we have made the substitution, $E_e \approx (2\pi m^3_e \nu/eB)^{1/2}$. From this exercise, we can see that a population of electrons with a power-law index, $p$, will produce a spectrum of synchrotron radiation that takes the approximate form of $dN_{\gamma}/d\nu \propto j(\nu)/\nu \propto \nu^{-(p+1)/2}$.

In the hot corona, we adopt a blackbody distribution to describe the X-rays, with a temperature of $T_X \sim 1-10 \, {\rm keV}$. TeV-scale gamma rays can collide with these X-rays to produce not only electron-positron pairs, but also muon-antimuon pairs. In the center-of-momentum frame, the total energy of such a collision is given by 
\begin{align}
E_{\rm CM} &= [2E_{\gamma} \epsilon_{X} \, (1-\cos \theta)]^{1/2} \\
&\approx 0.5 \, {\rm GeV} \times \bigg(\frac{E_{\gamma}}{10 \, {\rm TeV}}\bigg)^{1/2} \bigg(\frac{\epsilon_{X}}{27 \, {\rm keV}}\bigg)^{1/2} \bigg(\frac{1-\cos \theta}{0.5}\bigg)^{1/2},\nonumber
\end{align}
where $\theta$ is the angle between the incoming photons, and we have scaled the average energy of a target photon to the temperature of the hot corona, $\langle \epsilon_X \rangle \approx 2.7 \, T_X = 2.7-27 \, {\rm keV}$. Notice that for $E_{\gamma} \gsim {\rm TeV} \times (10 \, {\rm keV}/T_X)$, the energy of these collisions will typically exceed not only the threshold for electron-positron pair production, but also that for the production of muon-antimuon pairs.

\begin{figure}
\centering
\includegraphics[width=0.5\textwidth,clip=true]{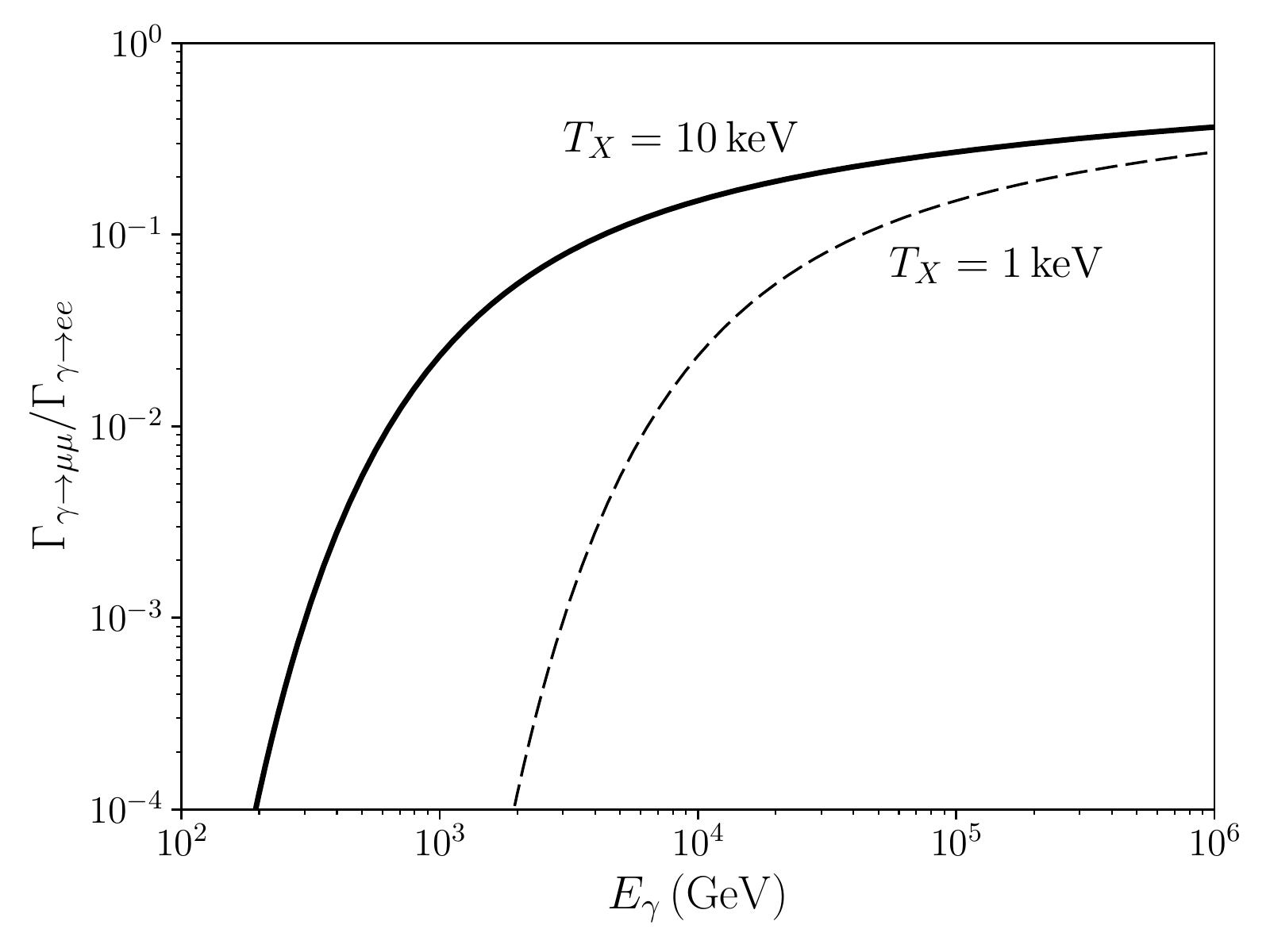} 
\caption{The ratio of the rates for muon-antimuon pair production and electron-positron pair production as a function of gamma-ray energy, for a blackbody spectrum of target photons with a temperature of $T_X=1 \, {\rm keV}$ (dashed) or $T_X=10 \, {\rm keV}$ (solid).}
\label{fig:muonfrac}
\end{figure}

\begin{figure*}
\centering
\includegraphics[width=0.49\textwidth,clip=true]{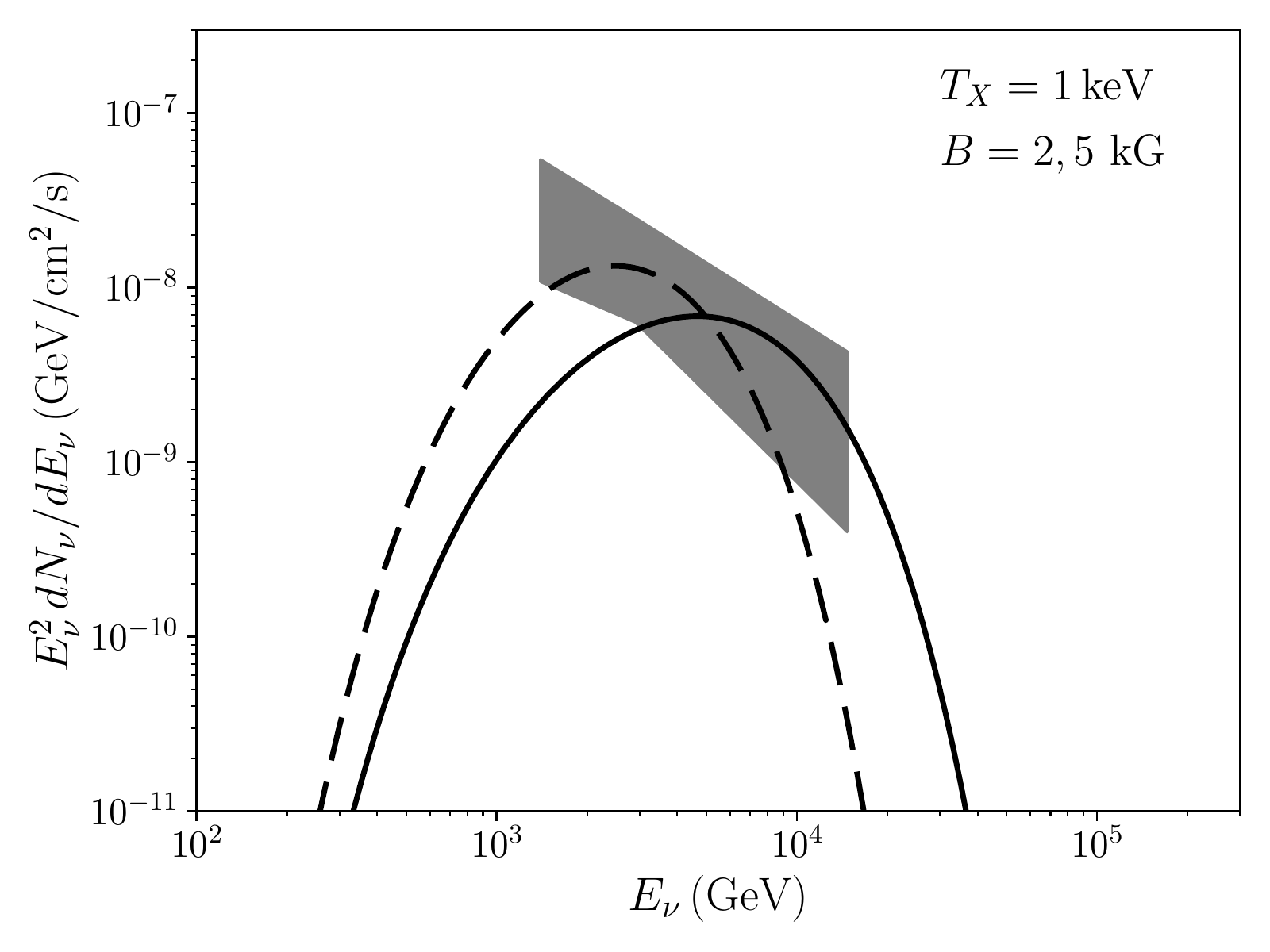} 
\includegraphics[width=0.49\textwidth,clip=true]{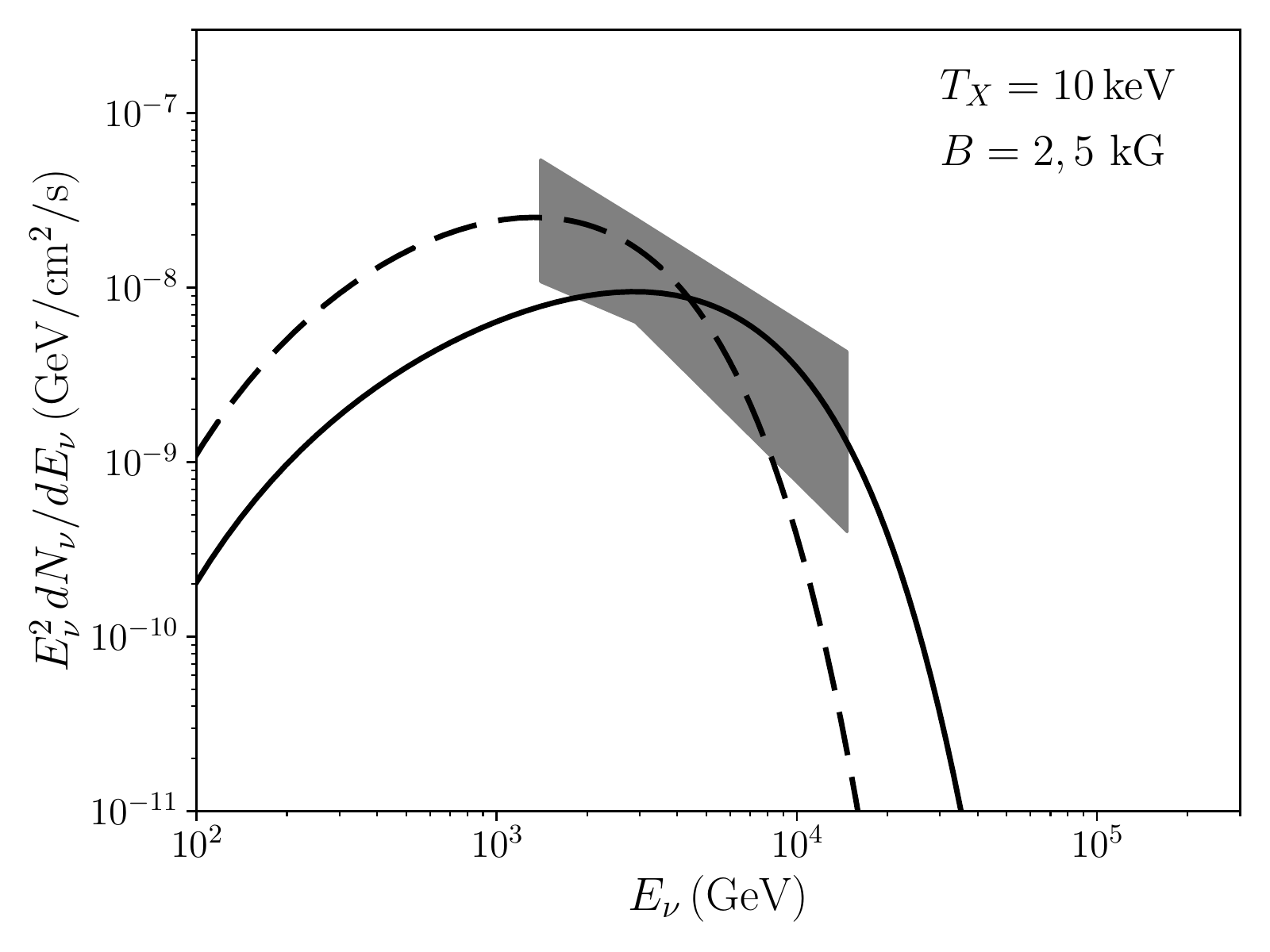} 
\caption{The all-flavor neutrino spectrum from muon-antimuon pair production in the core of NGC 1068, adopting a corona temperature of 1 keV (left frame) or 10 keV (right frame), magnetic fields of 2 or 5 kG (dashed and solid, respectively), and an electron spectrum given by $dN_e/dE_e \propto E_e^{-2} \, \exp(-E_e/300 \, {\rm TeV})$, normalized such that the total power injected above $E_e =1 \, {\rm TeV}$ is $4 \times 10^{42} \, {\rm erg/s}$ (left dashed), $1.2 \times 10^{43} \, {\rm erg/s}$ (left solid), $2 \times 10^{42} \, {\rm erg/s}$ (right dashed), or $6 \times 10^{42} \, {\rm erg/s}$ (right solid). The shaded band is the neutrino spectrum observed from NGC 1068, as reported by the IceCube Collaboration~\cite{IceCube:2022der}.}
\label{fig:neutrinospec}
\end{figure*}

The cross sections for electron-positron and muon-antimuon pair production are given by~\cite{Breit:1934zz,1983Afz....19..323A,Gould:1967zzb}:
\begin{align}
\label{eq:pairprod}
\sigma_{\gamma \gamma} = \frac{2 \pi \alpha^2}{E_{\rm CM}^2}  \bigg[2\beta (\beta^2-2)+(3-\beta^4) \ln\bigg(\frac{1+\beta}{1-\beta}\bigg)\bigg],
\end{align}
where 
\begin{align}
\beta &\equiv \bigg[1-\bigg(\frac{2m_{e, \mu}}{E_{\rm CM}}\bigg)^2\bigg]^{1/2}.
\end{align}
Using this cross section, we can integrate over the distribution of target photons to determine the rate of lepton-antilepton production for a gamma-ray propagating through an isotropic radiation field:
\begin{align}
\label{eq:eeopticaldepth}
\Gamma_{\gamma \gamma} =  \int^1_{-1}  \frac{1-\cos \theta}{2} \, d(\cos \theta) \int \sigma_{\gamma \gamma} \,\frac{dn_X}{d\epsilon_X} \, d\epsilon_X,
\end{align}
where $dn_X/d\epsilon_X$ is the differential number density of target photons. In Fig.~\ref{fig:muonfrac}, we show the ratio of the rates for muon and electron pair production as a function of gamma-ray energy. At very high energies, this curve asymptotically approaches unity, corresponding to equal rates for the production of electron-positron and muon-antimuon pairs. In the relativistic limit, these muons will carry half of the energy of the very high-energy photon, leading to neutrinos with $E_{\nu} \approx E_{\gamma}/6$.

We note that muon pair production will be further suppressed if keV-scale photons do not dominate the radiation fields that are present within the scattering region. In the presence of large number densities of $\sim \mathcal{O}({\rm eV -  keV})$ photons (see Ref.~\cite{Murase:2019vdl}), most of the very high-energy photons will instead scatter with this lower-energy radiation to produce electron-positron pairs.

\section{Neutrinos From Muon Pair Production in NGC 1068}

To calculate the neutrino spectrum from muon pair production in the core of an AGN, such as NGC 1068, we follow the procedure described in the previous section, adopting an electron spectrum of $dN_e/dE_e \propto E_e^{-2} \, \exp(-E_e/300 \,{\rm TeV})$.  The results of this calculation are shown in Fig.~\ref{fig:neutrinospec}, for several choices of the magnetic field strength and the temperature of the hot corona. In each case, we have normalized the electron spectrum such that the total power injected above 1 TeV is $\sim (2-12) \times 10^{42} \, {\rm erg/s}$, as indicated in the caption. This normalization has been set under the assumption that the corona is optically thick to very high-energy photons. If this is not the case, the normalization of the high-energy electrons would need to be increased accordingly. The shaded band in these figures represent the neutrino spectrum observed from NGC 1068, as reported by the IceCube Collaboration~\cite{IceCube:2022der}. We have treated this emission as isotropic, and have taken the distance to NGC 1068 to be 14.4 Mpc. Note that for the magnetic fields considered here, the timescale for muon synchrotron energy losses is much longer than the lifetime of these particles.

\section{Discussion and Summary}

In this letter, we have proposed a novel mechanism for the production of high-energy neutrinos in astrophysical environments which is purely leptonic and does not rely on the acceleration of protons or nuclei. Instead of generating neutrinos through the process of pion production, we suggest that very high-energy gamma rays could interact with X-rays in the source to produce muon-antimuon pairs, which subsequently decay to generate high-energy neutrinos. Such very high-energy photons could potentially be generated through either synchrotron or inverse Compton scattering, and could lead to a spectrum of TeV-scale neutrinos that is compatible with that recently reported from the active galaxy, NGC 1068.

We consider it unlikely that this mechanism is responsible for most of the diffuse neutrino spectrum reported by the IceCube Collaboration. In realistic astrophysical environments, muon pair production could efficiently produce neutrinos in the $\sim 1-100 \, {\rm TeV}$ range, but would not significantly contribute at higher energies. Although we have focused here on the cores of AGN, other environments in which high-energy photon are present in high-temperature radiation fields could also generate neutrinos through muon pair production. Gamma-ray bursts, for example, could be interesting in this context.

One way to potentially determine whether the neutrinos from a given source are produced through muon pair production or through pion production would be to measure their flavor ratios~\cite{Beacom:2003nh}. Whereas pion decay produces neutrinos in a ratio of $\nu_e:\nu_{\mu}:\nu_{\tau} = 1:2:0$ (which after oscillations becomes $\nu_e:\nu_{\mu}:\nu_{\tau} \approx 1:1:1$), muon decay yields $\nu_e:\nu_{\mu}:\nu_{\tau} = 1.5:1.5:0$ (or after oscillations, $\nu_e:\nu_{\mu}:\nu_{\tau} \approx 1.2:0.9:0.9$). Although such a measurement would certainly be very challenging~\cite{IceCube:2015rro}, this information could, in principle, be used to discriminate between these production mechanisms. We also point out that this mechanism can only produce neutrinos at energies above the threshold for muon pair production, corresponding to $E_{\nu} \gsim 0.3 \, {\rm TeV} \times (10 \, {\rm keV}/T) $. If the neutrino spectrum from NGC 1068 is observed to extend to lower energies, hadronic interpretations would be favored.


\bigskip
\bigskip

\begin{acknowledgments}

We would like to thank Takahiro Sudoh, Haocheng Zhang, Kohta Murase, Ke Fang, Tim Linden, Gordan Krnjaic, Carlos Blanco, Damiano Caprioli, and Rodolfo Capdevilla for helpful discussions. DH is supported by the Fermi Research Alliance, LLC under Contract No.~DE-AC02-07CH11359 with the U.S. Department of Energy, Office of Science, Office of High Energy Physics. KP is supported by the National Science Foundation under grant AST-1828784.

\end{acknowledgments}

\bibliography{muonpairhidden}

\clearpage
\onecolumngrid
\appendix
\setcounter{equation}{0}
\setcounter{figure}{0}
\renewcommand{\theequation}{S.\arabic{equation}}
\renewcommand{\thefigure}{S.\arabic{figure}}

\end{document}